\documentclass{optica-article}

\journal{opticajournal} 

\articletype{Research Article}

\usepackage{lineno}
\usepackage{float}

\begin{document}

\title{\emph{In situ} thermal trimming of waveguides in a standard active silicon photonics platform}

\author{Tianyuan Xue,\authormark{1,2,3} Hannes Wahn,\authormark{1} Andrei Stalmashonak,\authormark{1} Joyce K. S. Poon,\authormark{1,2} and Wesley D. Sacher\authormark{1,4}}

\address{
\authormark{1}Max Planck Institute of Microstructure Physics, Weinberg 2, Halle, Germany\\
\authormark{2}Department of Electrical and Computer Engineering, University of Toronto, 10 King’s College Rd., Toronto, Ontario M5S 3G4, Canada\\
\authormark{3}tianyuan.xue@mpi-halle.mpg.de\\
\authormark{4}wesley.sacher@mpi-halle.mpg.de
}


\begin{abstract*}  
We present suspended heater structures fabricated in a standard C- and O-band silicon (Si) photonics platform that can achieve sufficiently high local temperatures to induce effective refractive index trimming of Si and silicon nitride (SiN) waveguides with 30 - 90 mW of applied electrical power. Following thermal trimming at moderate powers ($\leq$ 40 mW), maximum changes in the averaged waveguide effective refractive index of $-5.18 \times 10^{-3}$ and $-7.9 \times 10^{-4}$ are demonstrated in SiN and Si waveguides, respectively, at a wavelength of 1550 nm. At higher powers, SiN waveguides exhibit positive averaged effective index changes up to $\approx$0.02, demonstrating bi-directional index trimming. As an example application, we demonstrate bias point trimming of a carrier injection Mach-Zehnder switch. Through investigations of the origin of the thermal trimming effect, we hypothesize that changes in the silica (SiO$_2$) waveguide cladding may be a primary underlying mechanism at temperatures $\gtrsim$300\textdegree C, with significant trimming of SiN waveguide cores occurring at larger temperatures $\gtrsim$510\textdegree C. As the trimming experiments represent a form of accelerated thermal aging, we estimate the aging behavior of the suspended heaters by fitting and extrapolating the measured datasets to 100 - 200\textdegree C operating temperatures over five years.
\end{abstract*}

\section{Introduction}

Silicon (Si) photonics leverages mature CMOS-compatible fabrication processes for dense co-integration of electronics and photonics, enabling an array of scalable microsystems solutions for applications including data/telecommunications \cite{Dong:12,Azadeh:15,Li:20}, light detection and ranging (LiDAR) \cite{Poulton:17,Zhang:22}, and quantum information \cite{Sibson:17,Zhang:19}. In the last decade, the integration of silicon nitride (SiN) waveguides in Si photonics platforms has become commonplace, owing to the better optical power handling, lower thermal sensitivity, and increased robustness to dimensional variations of SiN compared to Si waveguides \cite{Sacher2015}. 
Despite the maturity of Si photonics foundry fabrication, nanoscale variations in the dimensions of fabricated Si and SiN waveguide structures introduce optical phase errors between identically designed structures. These imperfections are detrimental to interferometric structures, such as Mach-Zehnder interferometers (MZIs) and ring resonators, where phase errors lead to variability in operating wavelengths and bias points. 
These device imperfections are typically overcome using additional phase tuners and power monitors to actively correct for phase errors, while also compensating for on-chip thermal fluctuations and locking devices to fluctuating laser wavelengths \cite{Lee:16,Dong:17}. 
Post-fabrication refractive index trimming can also compensate for fabrication-induced phase errors by permanently altering the refractive index of the waveguide materials in a controlled manner --- potentially reducing the variability in devices across Si photonic integrated circuits (PICs) and, correspondingly, reducing the required active tuning ranges. To this end, multiple methods of refractive index trimming have been previously demonstrated in a variety of photonic platforms. These methods include electron beam irradiation \cite{Schrauwen2008}, ion implantation \cite{Milosevic2018,Jayatilleka2021}, optical irradiation \cite{DePaoli2020,Guo2017,Biryukova2020,Haeiwa2004,Bachman2011,Lipka:14,belogolovskii2024}, and thermal treatment \cite{Xie2021,Spector2016,Milosevic2018,Jayatilleka2021,Hagan2019,belogolovskii2024}. Overall, these methods require additional fabrication and/or post-processing steps beyond standard Si photonics foundry fabrication processes, and generally, with the exception of Ref. \citenum{Jayatilleka2021}, are incompatible with trimming of packaged dies.

To overcome these limitations of trimming methods for Si photonic devices, we recently reported \emph{in situ} post-fabrication effective refractive index trimming of SiN waveguides for the visible spectrum using suspended heater structures \cite{Sacher2023,VIS_trimming_arxiv}. 
Here, we extend this method to near-infrared wavelengths and present thermal trimming of plasma enhanced chemical vapor deposited (PECVD) SiN waveguides using suspended heaters in a standard, active, SiN-Si-on-insulator (SOI) photonics platform for the C- and O-bands. The thermal isolation offered by suspended heaters enables semi-uniform temperature profiles within the heated regions --- allowing the SiN waveguides to reach high (300-400{\textdegree}C) trimming temperatures without excessive and potentially damaging temperatures in the resistive titanium nitride (TiN) heating elements (Section 2). In addition, the hot spots generated by suspended heaters are highly localized for independent trimming of multiple devices. The thermal isolation also enables trimming with only moderate (30 - 40 mW) electrical power dissipation, simplifying drive circuitry requirements. With this method, we demonstrate the operation of suspended heaters in two regimes where the effective index of the waveguides can be modified either transiently, as a thermo-optic phase shifter (at low applied electrical powers), or permanently, for the purpose of correcting fabrication-induced phase errors (at moderate applied powers), Section 3. Investigations of the origin of the trimming effects indicate that changes in the SiO$_2$ cladding of the waveguides with thermal treatment may be a primary mechanism. Following this hypothesis, we demonstrate effective index trimming of Si waveguides embedded in suspended heaters (Section 4). Additionally, for SiN waveguides trimmed at higher applied trimming powers, we observe a shift from negative to positive modal effective index changes --- demonstrating bi-directional trimming --- which we attribute to trimming of the SiN waveguide cores occurring at estimated temperatures $\gtrsim$510\textdegree C. In Section 5, extrapolations of the trimming data are used to estimate the thermal aging behavior of SiN waveguides in suspended heaters at 100 - 200\textdegree C operating temperatures. In these initial estimates, substantial drifts in optical phase are projected over multiple years at $\gtrsim$150\textdegree C. As the thermal trimming and aging behavior are expected to vary with the composition of the SiO$_2$ cladding, SiN cores, and passivation layer(s), further investigations are required to determine the variability in these characteristics across wafer lots and foundry processes.

Overall, the flexibility and simplicity of our trimming method, together with the maturity and widespread use of suspended heaters in Si photonics \cite{Fang:11,Lu:15,Celo:16}, opens an avenue toward practical trimming of large-scale Si photonic integrated circuits, potentially after packaging. In addition, the implications of this work toward the long-term thermal aging of waveguides are particularly relevant to the rapidly expanding use cases of dense Si PICs co-integrated with electronics operating at $\gtrsim$100\textdegree C (e.g., for high performance computing) \cite{Sahni:24,Kurata2021} --- with thermal trimming offering possibilities for mitigating (via device hardening) or counteracting (via phase error correction) these effects. An initial report of our results was presented in our conference abstract, Ref. \citenum{Xue2023}.

\section{Device design}

\begin{figure}[t!]
    \centering
    \includegraphics[width=1\textwidth]{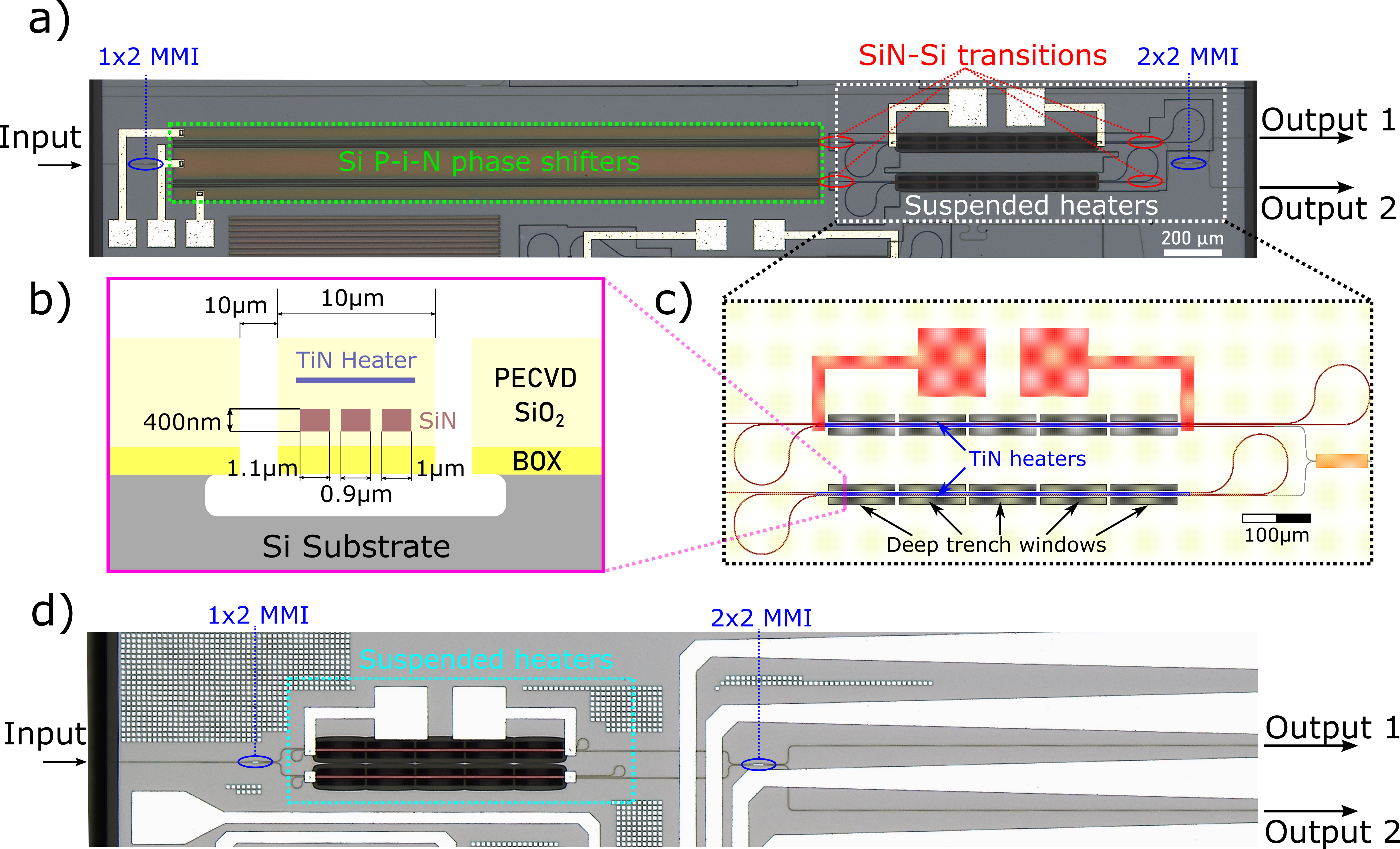}
            \caption{Trimming test devices. (a) Annotated optical micrograph of the SiN test device (imbalanced MZI switch including suspended heaters with embedded SiN waveguides, and also, Si PIN phase shifters). (b) Cross-section and (c) top-down schematics of the suspended heater structure. BOX: buried oxide. (d) Annotated optical micrograph of the Si test device, which is similar to (a) but with Si waveguides embedded in the suspended heaters and no PIN phase shifters. (a-c) Adapted from Ref. \citenum{Xue2023}.}
            \label{fig:circuit}
\end{figure}

To characterize the trimming method for SiN waveguides, we designed an imbalanced MZI test device, Fig. \ref{fig:circuit}(a). The MZI is formed by Si $1\times 2$ and $2\times 2$ multimode interferometers (MMIs) for splitting input light into two arms and subsequently interfering the paths at the outputs (Output 1 and Output 2). A suspended heater with folded PECVD SiN waveguides (with PECVD SiO$_2$ cladding) is integrated into one of the arms, with a nominally identical (but electrically disconnected) heater in the second arm for balancing optical losses. Each arm of the MZI test device also includes a Si PIN carrier injection phase shifter, enabling testing of trimmed MZI devices for fast optical switching (Section 3). The PIN phase shifter design is similar to Ref. \citenum{Green2007}, with a separation of 900 nm between the edges of the Si rib waveguide and the corresponding edges of the heavily-doped (p++ and n++) sections of the partially-etched Si slab. Adiabatic SiN-Si interlayer waveguide transitions \cite{Sacher2015} couple light between the Si and SiN waveguides before and after the suspended heater sections. To facilitate characterization of the thermally-trimmed phase shift, the MZI has a SiN waveguide length imbalance of 240 {\textmu}m between the arms. This imbalance generates fringes in the output transmission spectrum with a free spectral range (FSR) $\approx$ 5 nm around the target wavelength, $\lambda$, of 1550 nm.

A diagram of the suspended heater section is shown in Fig. \ref{fig:circuit}(c). The 10 {\textmu}m wide bridge is defined via a series of deep trench windows on either side, with 6 {\textmu}m wide anchors for structural support every 103 {\textmu}m along the bridge. Si undercut etching via the deep trench windows is used to release the bridge from the Si substrate, forming the suspended structure. The SiN waveguide is folded three times in this suspended bridge underneath a 510 {\textmu}m long TiN heater. Dissimilar waveguide widths of 1.1, 0.9, and 1 {\textmu}m are used for the three passes to limit inter-waveguide coupling while tightly packing the waveguides in the suspended bridge \cite{Lu:15}. Eigenmode expansion simulations (Ansys Lumerical) show a maximum coupling of -26 dB between the waveguides in the suspended bridge section.

\begin{figure}[t!]
    \centering
    \includegraphics[width=1\textwidth]{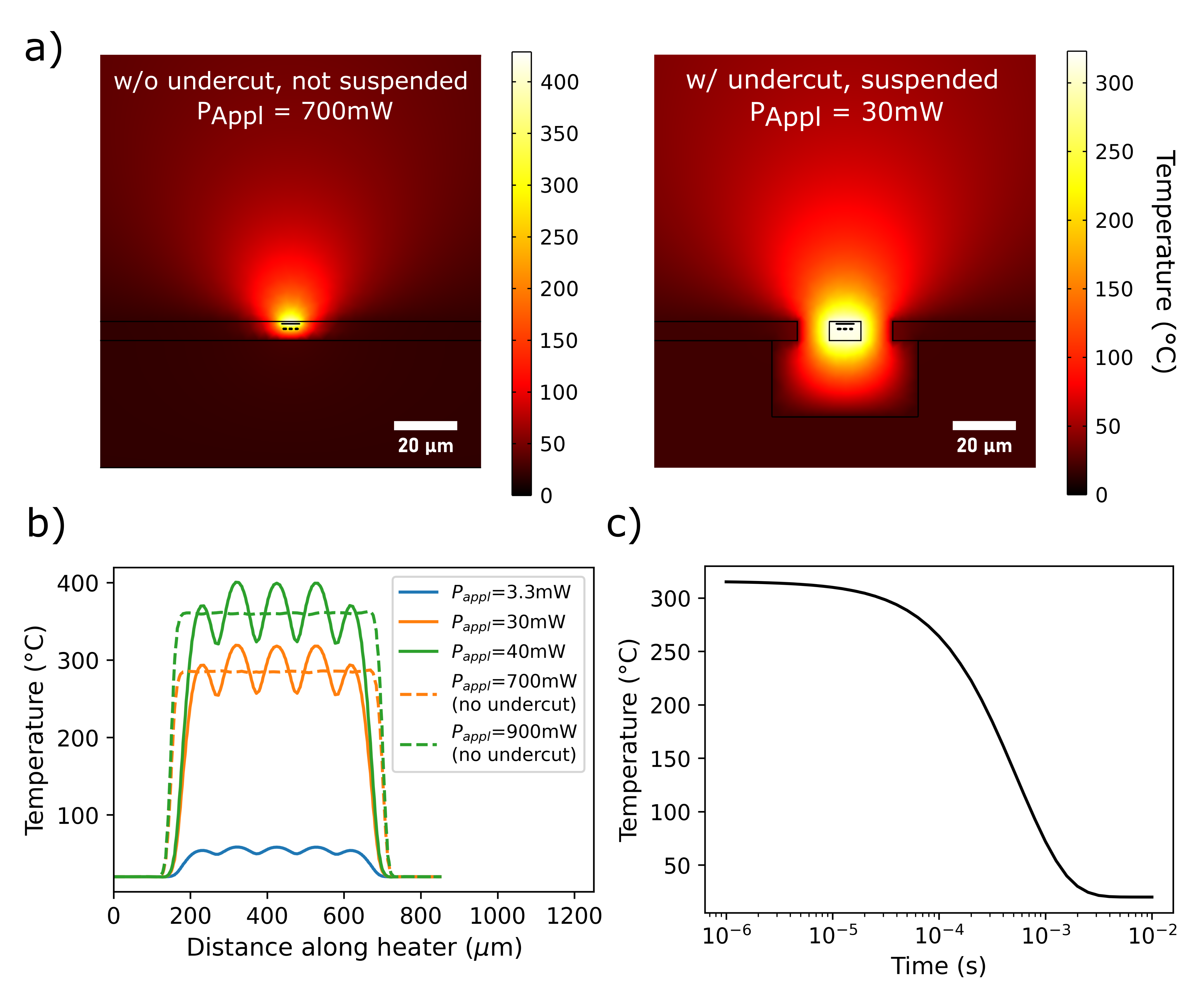}
            \caption{Thermal simulations of the SiN test device from COMSOL Multiphysics. (a) Cross-section thermal profiles of the heater structure without (left) and with (right) undercut etching of the Si substrate (to define a suspended structure). (b) Temperature profile along the center SiN waveguide for various applied electrical powers ($P_{Appl}$) with and without Si undercut. The ripples in the undercut curves are due to the periodic anchor structures of the suspended heater (for mechanical support). (c) Simulated cooling time of the SiN waveguides in the suspended heater with 30 mW of applied electrical power and an ambient temperature of 20{\textdegree}C}
            \label{fig:thermal}
\end{figure}

Figure \ref{fig:thermal} shows the simulated impact of the thermal isolation provided by suspending the heater region. Figures \ref{fig:thermal}(a) and \ref{fig:thermal}(b) show that, in the suspended structure, SiN waveguides are able to reach temperatures in excess of 300{\textdegree}C with only 30 mW of applied electrical power and a temperature difference of {5{\textdegree}C} between the TiN heater and SiN waveguides. 
Figure \ref{fig:thermal}(b) compares the temperature of the central SiN waveguide under the TiN heater at various applied electrical powers ($P_{appl}$) with and without suspension of the heater structure --- showing that $P_{appl} =$ 700 mW is needed to reach $\approx$300{\textdegree}C temperatures without the suspended structure. 
Heating the waveguides at that power without a suspended structure also results in a large temperature difference between the TiN and SiN of 146\textdegree C, which limits the maximum achievable temperature in the SiN waveguides before damaging the TiN heater. We also note that the abrupt temperature decrease outside the suspended heater presents the possibility for independent trimming of multiple adjacent devices, with the device density limited by thermal crosstalk. 

A second imbalanced MZI test device with Si waveguides in place of the SiN waveguides within the suspended heater structures was designed to investigate the origins of the trimming mechanism. 
The design is largely the same as the SiN device, but the Si PIN phase shifters are absent and the suspended heater region is slightly modified (6 \textmu m wide anchors for structural support every 95 \textmu m, for a total length of 469 \textmu m). 
The folded Si waveguide widths in the suspended heater region are 450, 550, and 500 nm. The Si waveguide path length imbalance between the MZI arms is 140 \textmu m to maintain a similar FSR as the SiN test device. 

The devices were fabricated in a standard, active, SiN-on-SOI photonics, multi-project 200-mm wafer run [CMC Microsystems -- Advanced Micro Foundry (AMF) silicon photonics process].

\section{SiN waveguide trimming characterization}

Chips with SiN test devices were mounted on a thermoelectric stage with a setpoint temperature of 21{\textdegree}C. Light from a tunable laser source (Keysight 81606A) was connected to a fiber polarization controller, and transverse electric (TE) polarized laser light was coupled onto/off the chips, Fig. \ref{fig:circuit}(a), through inverse taper Si edge couplers via lensed fibers. To determine the relative optical phase shift between the MZI arms, the transmission spectrum of Output 1 was measured with an optical power meter (Keysight N7745A) as the tunable laser wavelength was swept around $\lambda = 1550$ nm. The TiN heaters were driven by a sourcemeter (Keysight B2912A) via tungsten electrical probes contacting on-chip pads. 

The thermal trimming method was tested and characterized in the fabricated MZI devices by cycling through: (1) a trimming step, during which power was applied to the suspended heaters for a specified duration, (2) a cool-down period, wherein the heaters were turned off, and (3) a measurement of the optical transmission spectrum of Output 1. Each cooling period was $> 30$ s, which was longer than the simulated ($\sim$0.01 s) cooling period needed for the suspended heaters to return to ambient temperature, Fig. \ref{fig:thermal}(c). To maintain optimal and consistent transmission throughout the trimming tests, the input and output lensed fibers were periodically realigned to the on-chip edge couplers via piezoelectric micro-positioners.

The relative optical phase difference, $\phi$, at wavenumber $k_0$, between the two arms of the MZI can be inferred from the transmission spectrum according to
\begin{equation}
\phi(k_0) = k_0[n_{eff}(k_0)\Delta L + \Delta n_{eff}(k_0)L] \simeq 2\pi \left(\frac{k_0-k_1}{k_2-k_1}\right) + \phi_{offset},
\end{equation}
where $n_{eff}(k_0)$ is the effective index of the SiN waveguide at $k_0$, $k_1$ and $k_2$ are the wavenumbers corresponding to the troughs adjacent to $k_0$ in the measured spectrum, $\Delta L$ is the length imbalance between the two arms, $L$ is the length of the waveguide under the heater, and $\phi_{offset}$ is a phase offset determined by the selection of output port and phase shift sign convention. 
This method approximates the effective index of the SiN waveguide within a single FSR ($\approx 5$ nm) as constant [i.e., $n_{eff}(k_0) = n_{eff}(k_1) = n_{eff}(k_2)$]. The persistent change in the relative optical phase difference [$\Delta \phi (t)$] after a combined trimming time of $t$, is defined relative to the initial phase difference as
\begin{equation}
\Delta\phi(t) = \phi(t)-\phi(t=0).
\end{equation}

\begin{figure}[b!]
    \centering
    \includegraphics[width=1\textwidth]{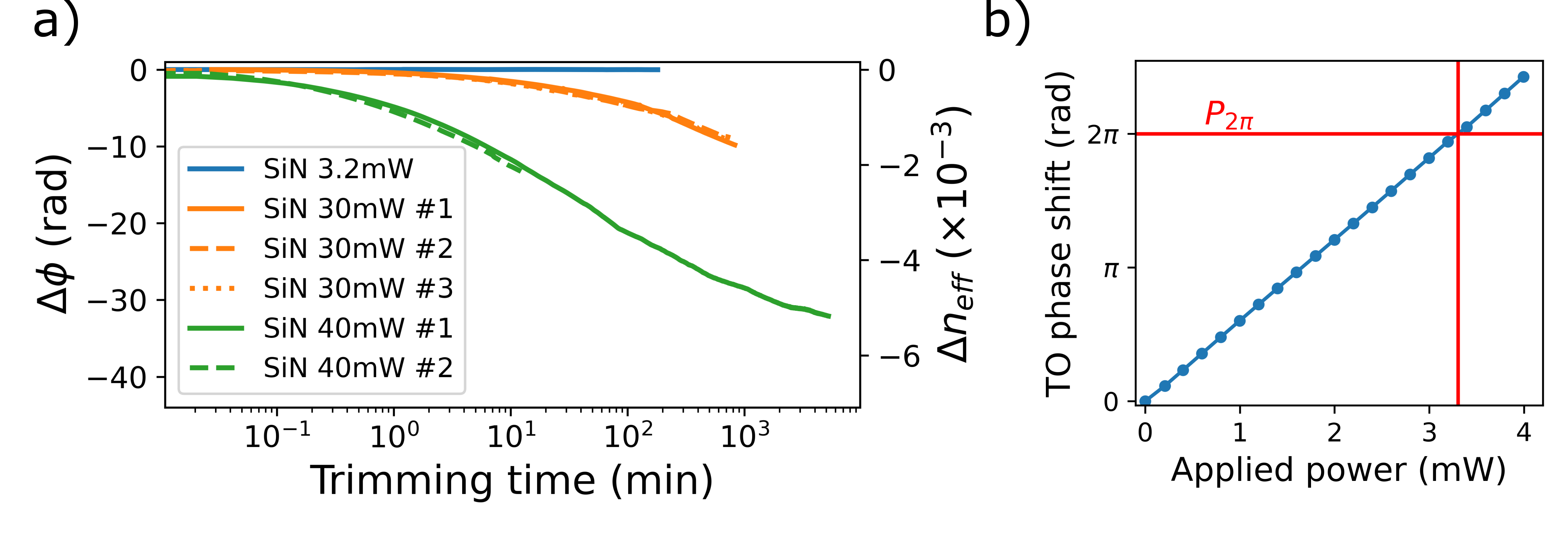}
            \caption{(a) Measured phase shift due to trimming ($\Delta \phi$) and the corresponding change in SiN waveguide effective index averaged over the suspended bridge section ($\Delta n_{eff}$) as a function of trimming time at applied electrical powers of 3.2 mW (one chip), 30 mW (three chips), and 40 mW (two chips). (b) Characterization of the thermo-optic (TO) phase shifter operating mode of a suspended heater.
            }
            \label{fig:trimming}
\end{figure}

The heaters were first characterized at low applied powers ($<5$ mW) as thermo-optic phase shifters. The measured thermo-optic response is shown in Fig. \ref{fig:trimming}(b) at $\lambda = $ 1550 nm. The applied electrical power for a $\pi$ radian phase shift, $P_{\pi}$, was 1.7 mW, and $P_{2\pi}$ was 3.3 mW. Higher applied electrical powers ($P_{Appl}$) were then used for thermal trimming. Figure \ref{fig:trimming}(a) shows the measured phase shift due to trimming and corresponding change in the effective index of the SiN waveguide under the heater for $P_{Appl} = $ 3.2, 30, and 40 mW at $\lambda = 1550$ nm. 
These different trimming conditions (each with different $P_{Appl}$) were applied to nominally-identical MZI test structures on different chips. 
2$\pi$ rad phase shifts were achieved in 111 s for $P_{Appl} = 40$ mW and 239 minutes for $P_{Appl} = 30$ mW, corresponding to $\Delta n_{eff} = 1.01 \times 10^{-3}$. After a longer trimming period of 5245 min at $P_{Appl} = 40$ mW, $\Delta n_{eff} = 5.18 \times 10^{-3}$ and the trimming appeared to approach saturation. 
At $P_{Appl} \approx P_{2\pi}$, there was negligible change in the effective index after 181.5 minutes of trimming ($|\Delta n_{eff}| < 4\times10^{-6}$) --- demonstrating that the suspended heaters can also act as thermo-optic phase shifters without hysteresis.

\begin{figure}[ht!]
    \centering
    \includegraphics[width=0.88\textwidth]{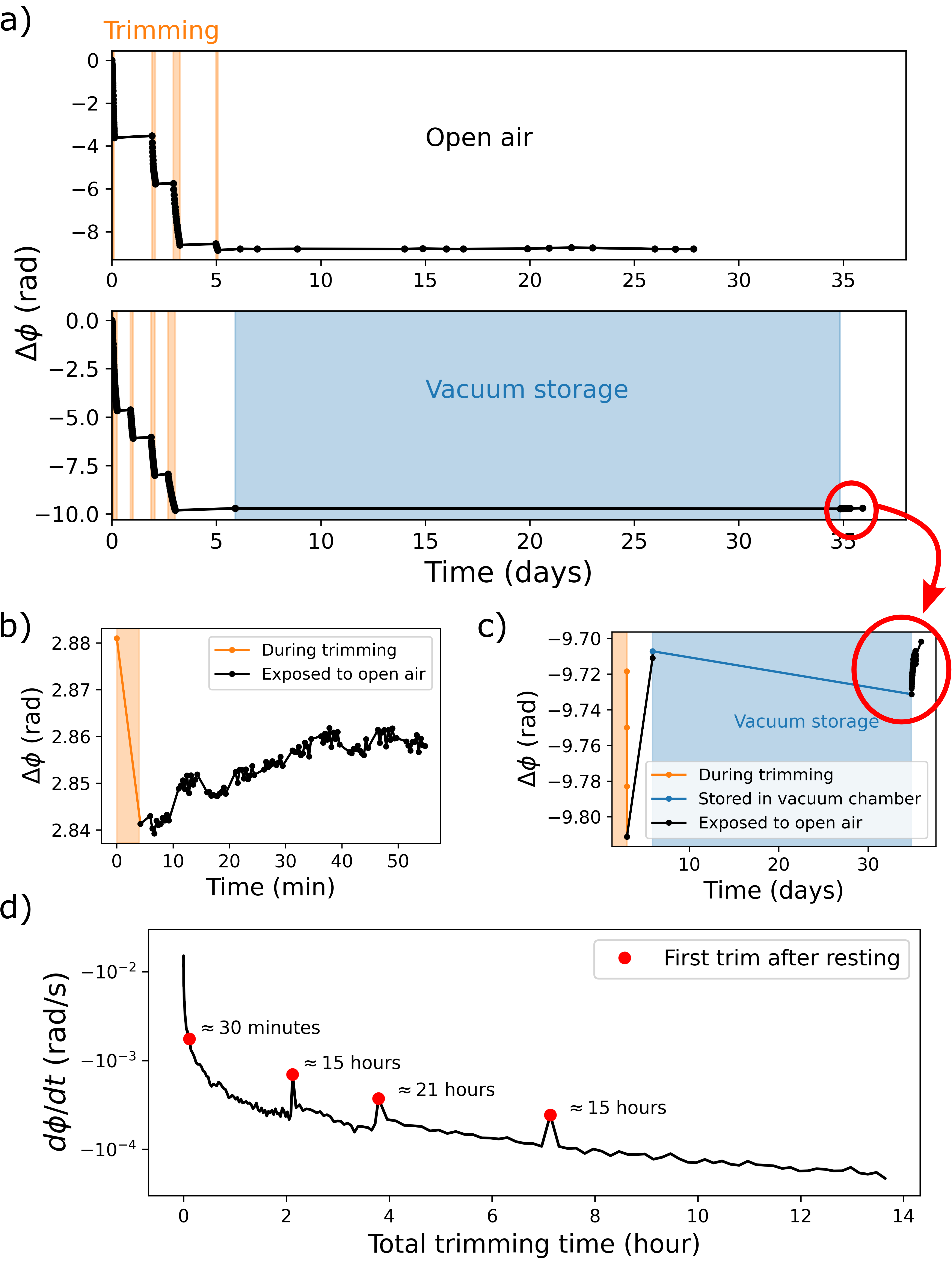}
            \caption{
            Long-term stability test for the trimmed phase shift of one sample in ambient conditions (21{\textdegree}C) (top) and another sample stored in a vacuum chamber (bottom); both samples were trimmed with 30 mW of applied power. The trimming periods are shaded in orange, and the period of time over which the sample was stored in a vacuum chamber is shaded in blue.
            (b,c) Observed recovery of the trimmed phase shift ($\Delta \phi$) after (b) a trimming period and (c) removal from storage in a vacuum chamber. (c) is a magnified view of the data in the bottom panel of (a).
            (d) Trimming rate ($d\phi/dt$) as a function of the trimming time with an applied electrical power of 30 mW. The red markers highlight the larger-than-expected trimming rate after an elongated period between trimming cycles ("resting period"). The resting periods are annotated for each red marker.}
            \label{fig:stability}
\end{figure}

Figure \ref{fig:stability}(a) shows the long-term stability of the trimmed phase shift. A power of 30 mW was applied to two samples: one was placed in a vacuum chamber and kept at a pressure of $\approx 7$ mbar at room temperature for 29 days, while the other remained on the characterization setup and was measured intermittently over the course of 23 days. 
Both samples showed a small drift in $\Delta \phi$ of $<$ 0.13 radians, equivalent to  $\Delta n_{eff} < 2 \times 10^{-5}$. 
As depicted in Figs. \ref{fig:stability}(b) and \ref{fig:stability}(c), some of this fluctuation can be attributed to a recovery process that was observed to occur in open air after either trimming or a period of vacuum storage. 
Furthermore, as shown in Fig. \ref{fig:stability}(d), the rate of change, d$\phi$/dt, is anomalously higher for the first trimming cycle following an extended (resting) period wherein no trimming was applied.
These larger-than-expected trimming rates resulted in an additional trimmed phase of 0.04 - 0.1 radians, similar in magnitude to the aforementioned drift in $\Delta \phi$.
A potential cause for these observations is the desiccation and rehydration of the PECVD SiO$_2$ cladding, similar to observations in Ref. \citenum{Wall2017}. 
From Fig. \ref{fig:stability}(d), this reversible process appears to be independent from the bulk of the trimmed phase, which did not exhibit recovery over time. 
Overall, the SiN waveguide effective refractive index change with thermal trimming was observed to be stable over the monitoring period of 23 days in air and 29 days in vacuum at room temperature.

\begin{figure}[t!]
    \centering
    \includegraphics[width=0.93\textwidth]{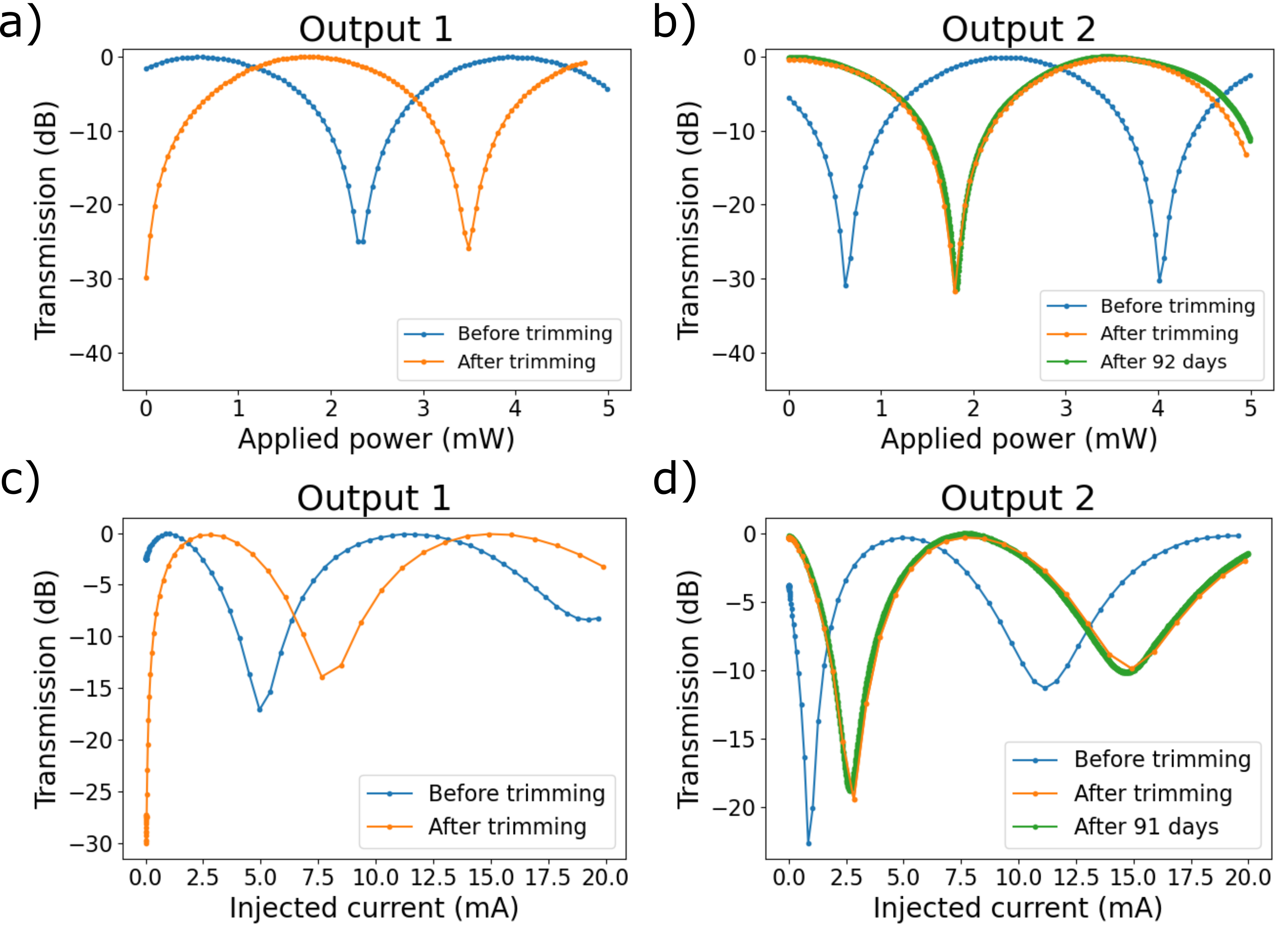}
            \caption{Demonstration of thermal trimming for bias power reduction of an MZI switch (SiN test device). Transmission of the MZI measured at (a) Output 1 and (b) Output 2 vs. applied electrical power to the suspended heater before trimming, after trimming, and after 92 days (Output 2 only) at $\lambda =$ 1550 nm. Transmission measured at (c) Output 1 and (d) Output 2 vs. current injected into the PIN phase shifter before trimming, after trimming, and after 91 days (Output 2 only) at $\lambda =$ 1550 nm.}
            \label{fig:bias}
\end{figure}

\begin{figure}[ht!]
    \centering
    \includegraphics[width=0.93\textwidth]{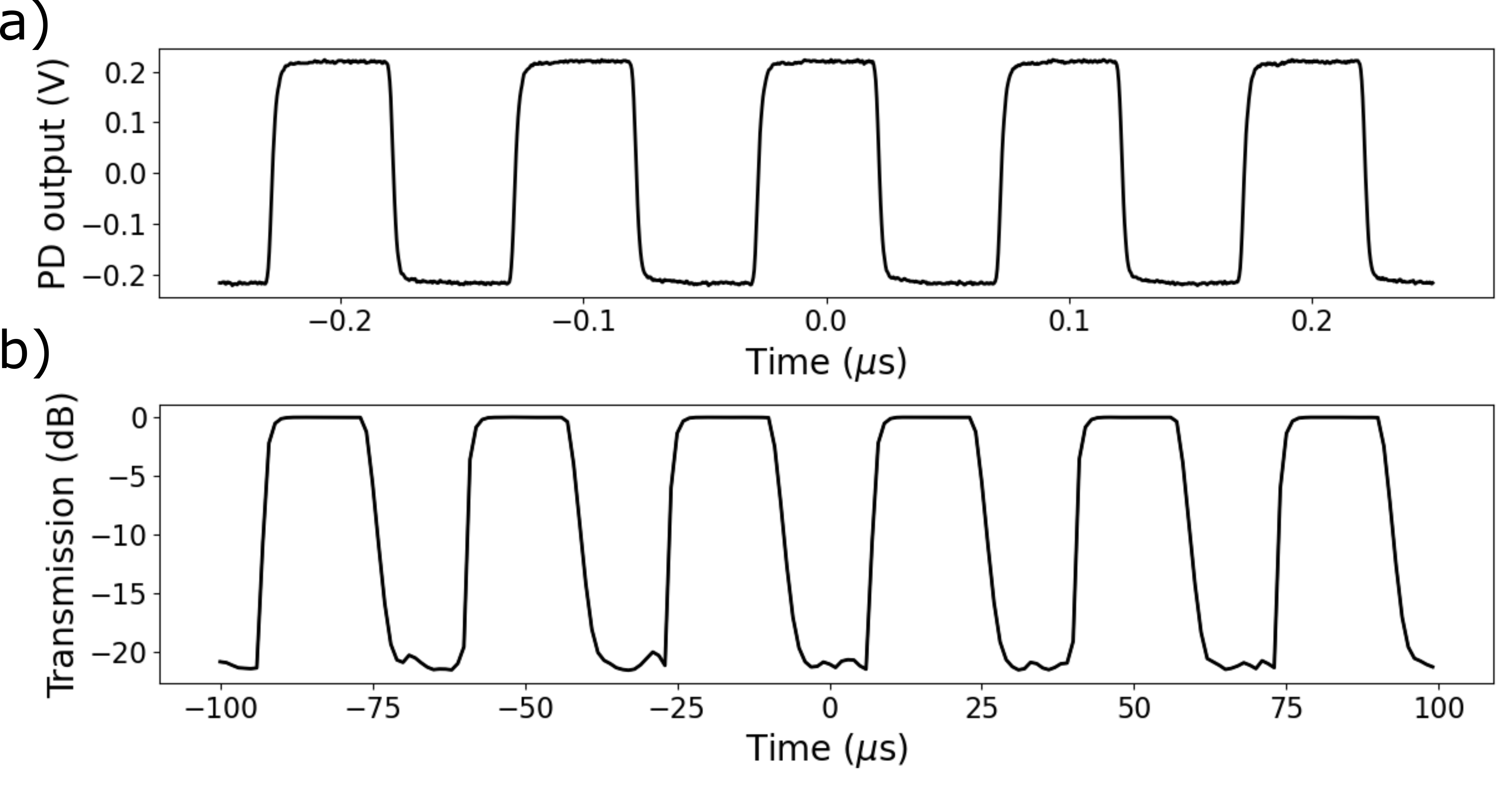}
            \caption{(a) Modulated transmission of the SiN test device MZI (Output 1) with the PIN phase shifter driven at 10 MHz. An AC-coupled photodetector (PD) was used for the measurement. (b) Modulated transmission at 30 kHz measured by an optical power meter}
            \label{fig:transient}
\end{figure}

The above-mentioned small recovery in the trimmed phase shift is predictable and can be taken into account during the trimming process. 
To demonstrate this, we trimmed one of the SiN test device MZIs to a target bias point of $\phi = -\pi/2$ (at $\lambda = $ 1550 nm) by over trimming (beyond the desired bias point) to compensate for the recovery effect. 
The process of controlled trimming of the MZI is detailed in Appendix A. The results are summarized in Fig. \ref{fig:bias}, and the final measured phase difference is within 0.013 rad (0.74{\textdegree}) of the targeted bias point of $\phi = \frac{-\pi}{2}$. 
Figures \ref{fig:bias}(a)-(d) show the transmission of Outputs 1 and 2 as a function of the applied power to the thermo-optic phase shifter and injected current to one of the PIN phase shifters before and after trimming of the MZI. Figures \ref{fig:bias}(b) and \ref{fig:bias}(d) also show the transmission of Output 2 after 91-92 days of storage. 
This measurement was not performed for Output 1 due to damage to the corresponding edge coupler during measurements. Figure \ref{fig:transient}(a) shows the operation of the trimmed MZI as a fast optical switch using the PIN carrier-injection phase shifter operating at 10 MHz without any applied bias signal. Figure \ref{fig:transient}(b) shows the transmission at Output 2 of the optical switch with the PIN phase shifter driven at 30 kHz, with > 20 dB of optical extinction ratio between the on and off states. 
To achieve the same bias point via active tuning rather than trimming, the PIN phase shifter would consume 4.6 mW of excess power and the heater would consume 2.3 mW. These excess power figures depend on the initial phase difference between the MZI arms (prior to trimming, $\phi =$ 3.632 rad in Fig. \ref{fig:bias}). A histogram of the initial phase difference for nominally-identical MZIs across 12 chips is shown in Appendix C; $\phi$ spans 1.94 - 5.96 rad.

\section{Investigation of trimming mechanism and Si waveguide trimming}
\begin{figure}[hb!]
    \centering
    \includegraphics[width=1\textwidth]{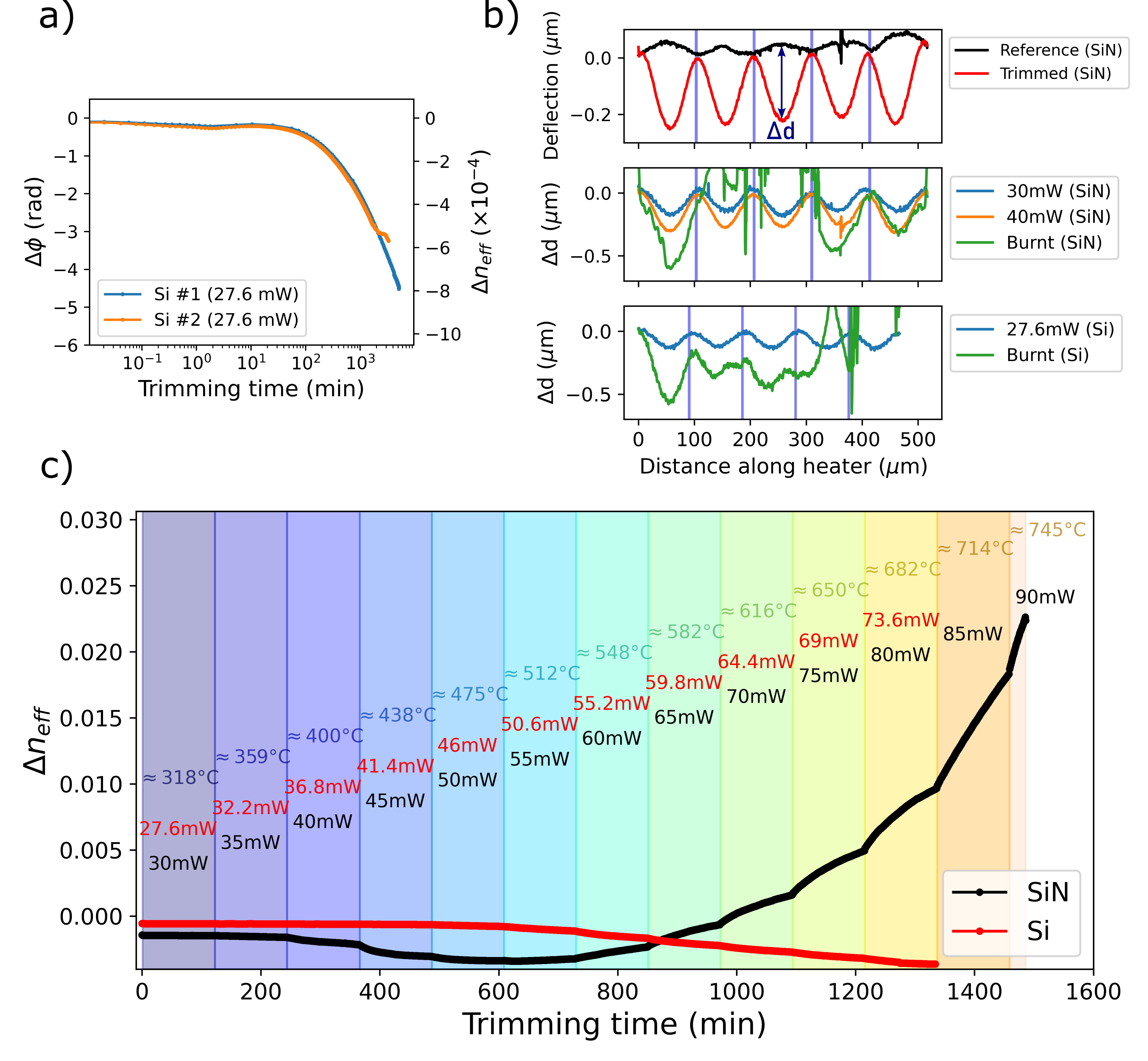}
            \caption{Investigation of trimming mechanisms. 
            (a) Trimming of Si waveguides. Phase shift due to trimming ($\Delta \phi$) and the corresponding change in Si waveguide effective index averaged over the suspended bridge ($\Delta n_{eff}$) vs. trimming time with $P_{Appl} =$ 27.6 mW. 
            (b) Surface profiles of suspended heaters with embedded SiN and Si waveguides; the Reference suspended heater was in the opposite (unused) MZI arm to the trimmed arm. Deflection of Reference and trimmed ($P_{Appl} =$ 40 mW) SiN suspended heaters (top). Change in deflection ($\Delta d$) between the Reference and trimmed SiN suspended heater driven with $P_{Appl} =$ 30 mW, 40 mW, and until failure (burnt) (middle). $\Delta d$ between the Reference and trimmed Si suspended heater driven at $P_{Appl} =$ 27.6 mW and until failure (bottom). Blue vertical lines denote the locations of the anchors where the suspended structure is supported. 
            (c) $\Delta n_{eff}$ of SiN and Si waveguides in suspended heaters driven at progressively higher $P_{Appl}$ (120 min per step) until failure. The simulated peak temperature of the SiN and Si waveguides are shown at the top. The Si suspended heater power was stepped prematurely to 85 mW after 73.6 mW of applied power leading to device failure before measurements were performed. The non-zero starting point of each trimming curve is representative of the index change resulting from prior trimming tests on each sample.}
            \label{fig:burn}
\end{figure}

To isolate the origins of the trimming effect, an additional imbalanced MZI with Si (rather than SiN) waveguides in the suspended heater, Fig. \ref{fig:circuit}(d), was designed and fabricated. 
This test device underwent the same trimming protocol as the suspended heaters with SiN waveguides. Accounting for differences in the suspended structure dimensions between the two devices, a slightly lower applied electrical power was used for the Si device to reproduce similar peak temperatures to the SiN device (27.6 mW instead of 30 mW, based on simulations). 
Investigating two samples, trimming of the Si waveguides was observed with the same sign of $\Delta n_{eff}$ as the SiN waveguide trimming tests, Fig. \ref{fig:burn}(a). $\Delta n_{eff} = -3.3 \times 10^{-4}$ after 800 min, which was about 5 times lower in magnitude than the measured SiN waveguide trimming (at 30 mW), Fig. \ref{fig:trimming}(a). 
With the exception of a small dip in trimmed phase and index in the first 10 minutes of trimming, the Si and SiN trimming curves were similar in shape. 
Since the refractive index of intrinsic Si waveguide cores is unlikely to be permanently altered by the trimming temperatures in this work, these observations are indicative of the trimming effects originating from the common features shared by the Si and SiN devices --- the SiO$_2$ cladding and suspended structure. 
The differences in trimming magnitude between Si and SiN devices may be explained in part by the differing modal overlap with the PECVD SiO$_2$ cladding (12.4\% and 37.9\% for Si and SiN waveguides, respectively) and thermally grown buried oxide (BOX) (10.6\% and 1.3\% for Si and SiN waveguides). The PECVD SiO$_2$ and BOX may also have different trimming characteristics.

To further investigate the underlying trimming mechanism, surface profile measurements of the trimmed suspended heaters and the unused suspended heater (in the other arm of each MZI) were measured with a laser scanning confocal microscope (Keyence VK-X3000) and compared in Fig. \ref{fig:burn}(b). 
With increasing trimming temperature, the suspended structures tended to bow increasingly downward (toward the substrate). 
We hypothesize that this is a consequence of stress changes in the PECVD SiO$_2$ cladding of the waveguides due to hydrogen release from decomposition of SiOH bonds in the SiO$_2$, a well-documented phenomenon \cite{Ghaderi2016,Fu2017,Bigl2015}. 
Meanwhile, the BOX component of the cladding is expected to have a much lower hydrogen content (and contribution to these effects) due to the high growth temperature relative to PECVD SiO$_2$ \cite{Beckmann1971,Pan1985,Ermolieff1993}. 
This hypothesis is supported by our recent work in Ref. \citenum{VIS_trimming_arxiv}, where we demonstrated thermal trimming of SiN waveguides at visible wavelengths in a short-wavelength integrated photonics platform (also fabricated at Advanced Micro Foundry). 
Fourier-transform infrared (FTIR) spectroscopy measurements of the SiN and PECVD SiO$_2$ cladding (before and after thermal treatment at 350 - 450\textdegree C) revealed little to no change in the peaks associated with the SiN, but a significant change in the SiOH peak associated with the PECVD SiO$_2$. 
From these two observations, we further hypothesize that the modal effective index trimming demonstrated here stems from composition (and corresponding index) changes in the PECVD SiO$_2$ waveguide cladding. 
Prior studies of thin film SiO$_2$ thermal annealing report reductions in refractive index for temperatures < 600\textdegree C \cite{Adams1981,Fitch1990,Haque1995,Ghaderi2016} --- consistent with our observations. 
The resulting strain in the suspended structure may also contribute to the index changes. 

As the last component to our analysis of the trimming mechanism, we drove suspended heaters in the SiN and Si devices to progressively higher temperatures until the TiN heater failed. 
As shown in Fig. \ref{fig:burn}(c), while the Si waveguide was consistently trimmed in the same direction ($\Delta n_{eff} < 0$), the SiN waveguide trimming reversed direction ($\Delta n_{eff} > 0$) for $P_{Appl} > 55$ mW with an overall net positive $\Delta n_{eff}$ before failing. 
Simulations indicate that this transition in SiN waveguide trimming behavior occurred at a peak temperature of $\approx$ 512\textdegree C, similar to reported temperatures for thermal decomposition of Si-H and N-H bonds in PECVD SiN \cite{Lavareda2023}. 
Meanwhile, as shown in Fig. \ref{fig:burn}(b), the deflection of the SiN suspended heater suggests no remarkable change in trimming behavior of the SiO$_2$ cladding above this 55 mW threshold (continued downward bowing, with the exception of damaged / "burnt" sections of the bridge).  
The properties of the SiO$_2$ cladding (as the bulk of the suspended structure) are expected to dominate the deflection, and when the SiN device was driven to failure (beyond the 55 mW threshold), the suspended heater deflection was consistent with trends observed at lower applied powers and with its Si counterpart --- despite the reversal in effective modal index change. 
Together, these observations indicate that trimming of the SiN (waveguide core) index is responsible for the change of direction in $\Delta n_{eff}$ for $P_{Appl} > 55$ mW.

In summary, our hypothesis of the trimming mechanisms is as follows. (1) Moderate $P_{Appl} = $ 30 - 40 mW ($\approx$ 300 - 400\textdegree C) changes the index of the PECVD SiO$_2$ cladding due to decomposition of SiOH bonds, resulting in $\Delta n_{eff} < 0$ for both SiN and Si waveguides. 
(2) Higher $P_{Appl} >$ 55 mW ($>$ 510\textdegree C) changes the index of PECVD SiN due to decomposition of Si-H and/or N-H bonds, corresponding to $\Delta n_{eff} > 0$ for SiN waveguides. (3) $P_{Appl} >$ 55 mW also continues to change the index of the PECVD SiO$_2$ cladding, which leads to continued trimming with $\Delta n_{eff} < 0$ for Si waveguides. However, for SiN waveguides, $|\Delta n_{eff}|$ from cladding index changes is lower than the contribution from the SiN waveguide core. 

This hypothesis points to potential routes for device improvements and future investigations. 
First, the trimmed $|\Delta n_{eff}|$ of SiN and Si waveguides at moderate applied powers may be increased by maximizing overlap with the PECVD SiO$_2$ cladding (e.g., by reducing the waveguide width and/or thickness). 
Second, for increased functionality and flexibility, SiN waveguides can be trimmed with either $\Delta n_{eff} < 0$ or $> 0$ depending on the applied power, and engineering of the modal overlap with SiN vs. SiO$_2$ may define $|\Delta n_{eff}|$ in each operating mode. 
Alternatively, two cascaded suspended heaters may be used: one optimized for positive $\Delta n_{eff}$ (high SiN modal overlap) and the other for negative $\Delta n_{eff}$ (high SiO$_2$ modal overlap).

\section{Implications for long-term thermally induced device aging}
The thermal treatment of waveguides in suspended heaters is a proxy for accelerated aging \cite{Erdogan1994,Williams1995,Nemilov2000}, and the observed decrease in the effective refractive index of waveguides with thermal trimming is also expected to occur over longer periods of time at lower temperatures. In the following analysis, we apply the demarcation energy approximation \cite{Erdogan1994} to estimate the thermal aging behavior of SiN waveguides embedded within suspended heaters over multiple years at 100-200\textdegree C. These operating temperatures are relevant to dense PICs co-integrated with electronics for data communication and high performance computing \cite{Sahni:24,Kurata2021}, with thermo-optic phase shifters generating localized hot spots atop $\gtrsim$100\textdegree C baseline PIC temperatures. Optical phase shifts stemming from thermally induced aging may introduce design considerations for the active tuning ranges of Si photonic devices (to compensate aging) and their maximum operating temperatures (to limit aging).

The demarcation energy ($E_d$) is used to approximate the progressive annealing of a population of defects distributed over a range of activation energies, $E_a$. Defects with $E_a < E_d$ are assumed to be completely annealed, while defects with $E_a > E_d$ remain. 
The demarcation energy is defined as a function of the anneal temperature ($T$) and time ($t$):
\begin{equation}\label{eq:Ed}
E_d = k_BTln(\nu_0t),
\end{equation}
where $k_B$ is the Boltzmann constant and $\nu_0$ is the attempt frequency.
As the temperature varies throughout the suspended section due to the anchor structures, two sets of calculations were performed --- where (1) the maximum and (2) the minimum simulated temperature throughout the suspended structure (at each applied electrical power) was used to calculate the demarcation energy.
We then followed a similar procedure as our prior work in Ref. \citenum{VIS_trimming_arxiv}. We aligned the measured SiN waveguide trimming curves [Fig. \ref{fig:trimming}(a)] onto a trimmed phase shift versus $E_d$ curve using Eq. \ref{eq:Ed} with tuning of $\nu_0$. Subsequently, a master curve was fit to the aligned measurement data assuming the defects follow a Gaussian distribution of activation energies, $g(E)$.

\begin{figure}[]
    \centering
    \includegraphics[width=0.93\textwidth]{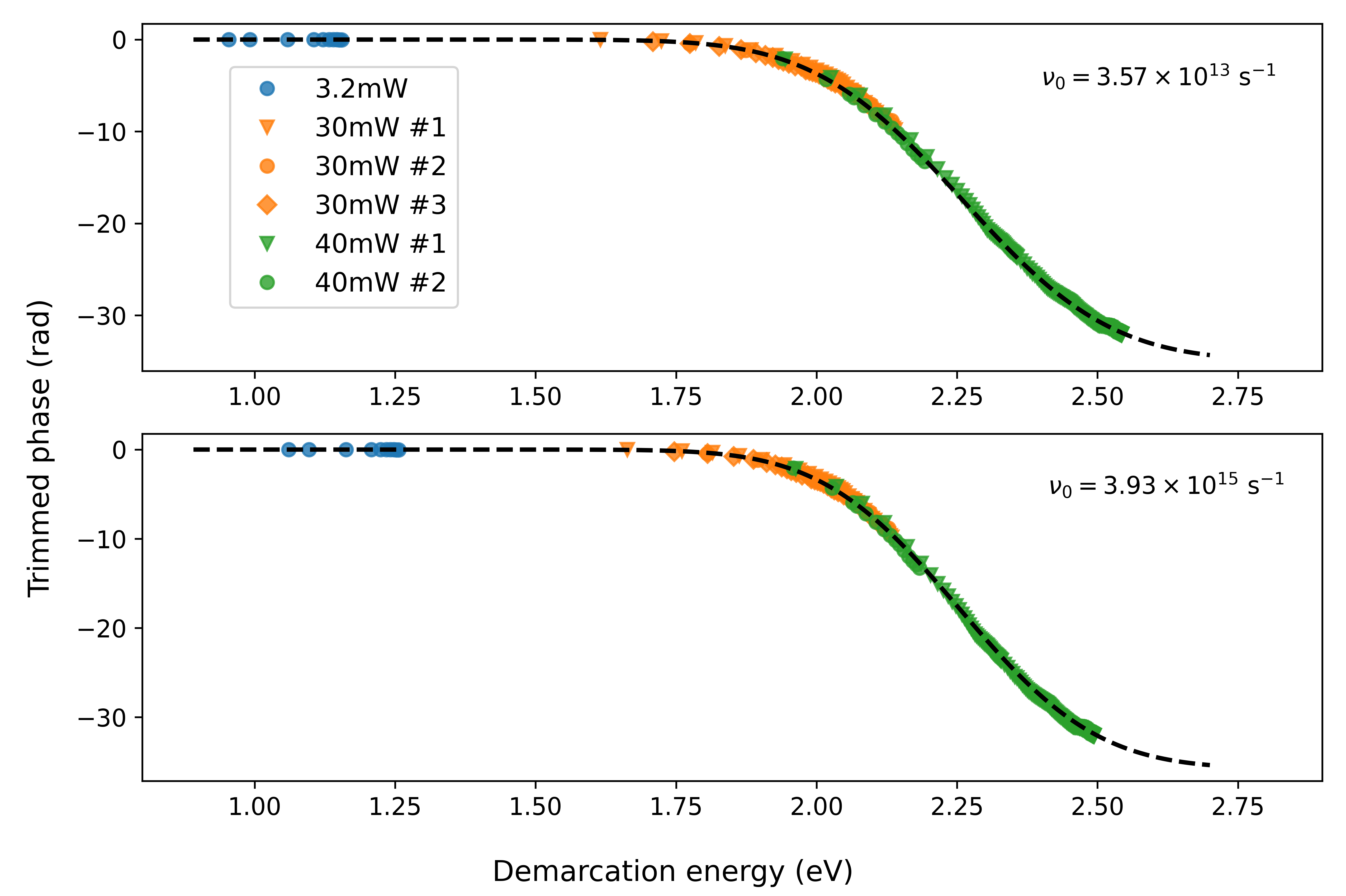}
            \caption{Plots of trimmed phase as a function of the demarcation energy at applied electrical powers of 3.2 mW, 30 mW, and 40 mW. (Top) Peak simulated temperatures in the suspended structure are assumed throughout the heater, corresponding to $\nu_0 = 3.57 \times 10^{13}$ s$^{-1}$. (Bottom) Lowest simulated temperatures in the suspended structure are assumed throughout the heater, corresponding to $\nu_0 = 3.93 \times 10^{15}$ s$^{-1}$. Black dashed curves: master curves fit to the aligned trimming datasets with the assumption of an underlying Gaussian distribution of activation energies for thermally annealed defects. The markers are under sampled from the experimental data for improved visibility.}
            \label{fig:Ed}
\end{figure}

\begin{figure}[]
    \centering
    \includegraphics[width=0.85\textwidth]{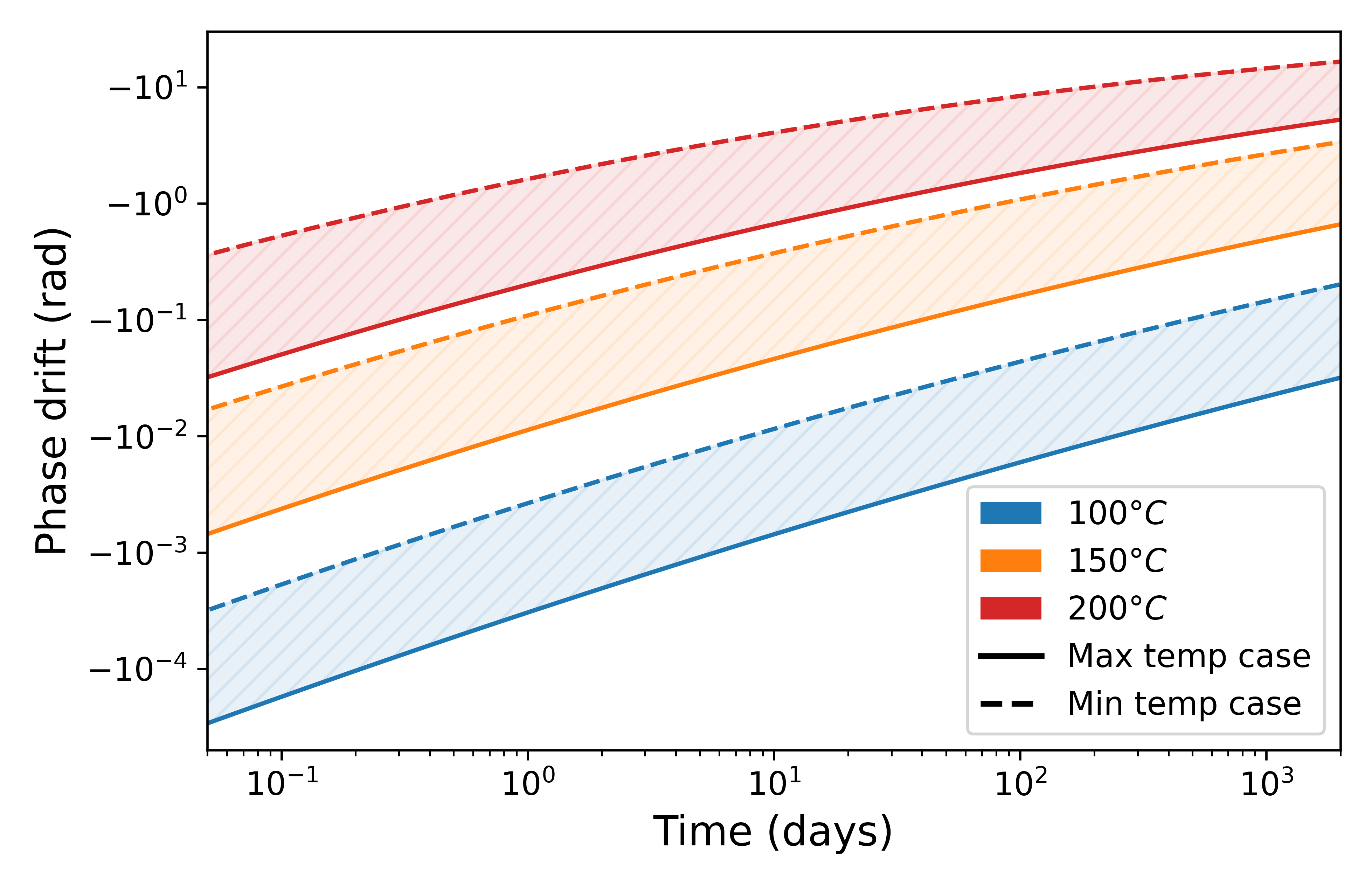}
            \caption{Estimated phase drift of the suspended TO phase shifter with SiN waveguides over five years at operating temperatures of 100, 150 and 200\textdegree C. Solid (dashed) curves: peak (lowest) simulated temperatures in the TO phase shifter are assumed throughout the suspended structure.}
            \label{fig:aging}
\end{figure}

The results are summarized in Fig. \ref{fig:Ed}.
$\nu_0$ was found to be $3.57\times10^{13}$ s$^{-1}$ and $3.93\times10^{15}$ s$^{-1}$ for the maximum and minimum temperature cases, respectively. Also, in the maximum (minimum) temperature cases, the fitted $g(E)$ had a mean activation energy of 2.26 eV (2.26 eV) and full width at half maximum of 0.49 eV (0.46 eV).
Using these two scenarios, we draw bounds for the estimated phase drift due to thermal aging of the suspended thermo-optic phase shifter with SiN waveguides, Fig. \ref{fig:aging}. At an operating temperature of 100\textdegree C, the phase drift in the maximum (minimum) temperature cases is -0.01 (-0.09) to -0.03 (-0.19) rad. for 1 to 5 years of exposure. Larger phase drifts are estimated at higher temperatures, and at 150\textdegree C, the phase drift is -0.31 (-1.83) to -0.64 (-3.27) rad. after 1 to 5 years. At 200\textdegree C, the phase drifts by -2.98 (-11.64) to -5.10 (-16.26) rad. over the same time periods. 

The accuracy of these thermal aging estimates is limited by the non-uniformity of the temperature profile, the accuracy of the simulated temperatures, and the assumption of $g(E)$ as a Gaussian distribution. We note that, due to the exponential dependence of the Boltzmann factor on temperature, the segments of the suspended structure at higher temperatures are expected to contribute significantly more to the observed phase trimming compared to the segments at lower temperatures --- suggesting that the maximum temperature case (lower bound of the estimated phase drift magnitude) should be the better approximation.
Furthermore, the attempt frequency needed to construct the master curve in the minimum temperature case is significantly higher than expected for annealing in solids, whereas the maximum temperature case is within the expected range (order of $10^{13}$ s$^{-1}$) \cite{Corbett1966}. Despite the limitations of these initial estimates of thermally induced aging, the above analysis indicates the potential for large phase drifts ($\gtrsim$ 1 rad.) after multiple years at operating temperatures of 150-200\textdegree C. Improvements to the aging estimates are expected from design modifications to the heaters for increased temperature uniformity (possibly by removing the anchors) in addition to direct temperature measurements of the suspended structures. Moreover, since the trimming and aging characteristics of the waveguides are intrinsically linked to the material properties of the cladding, core, and, potentially, passivation layer(s), future studies are needed to quantify the wafer-to-wafer and foundry-to-foundry dependence of these effects.  

As we discussed in Ref. \citenum{VIS_trimming_arxiv}, thermal trimming offers two avenues for mitigating thermal aging effects. First, thermal trimming may be used to correct for phase drifts due to aging. Leveraging the compatibility of our trimming method with packaged PICs, such corrections may be performed periodically over a multi-year span. Higher temperature trimming ($\gtrsim 510$\textdegree C) of SiN waveguides ($\Delta n_{eff} > 0$) may directly counteract the negative modal effective index changes due to aging. Trimming of Si waveguides or lower temperature trimming ($\approx$ 300-400\textdegree C) of SiN waveguides would require a 2$\pi$ rad. offset to counteract phase drifts with thermal aging (as both the trimming and aging exhibit $\Delta n_{eff} < 0$). Second, noting the saturation of $\Delta n_{eff}$ with trimming [Fig. \ref{fig:trimming}(a)] and aging (Fig. \ref{fig:aging}) time, thermal treatment may be used to harden waveguides against thermal aging. Sensitive photonic structures such as microring resonators and folded waveguides in interferometers may be embedded in suspended heaters and exposed to an initial extended period of thermal treatment --- depleting the defects which contribute to phase drifts with thermal aging.

\section{Conclusion}

In summary, we have demonstrated \emph{in situ} thermal trimming of waveguides using suspended heater structures in a standard active SiN-on-SOI photonic platform for the C- and O-bands. The thermal isolation of suspended heaters enables sufficient temperatures in SiN and Si waveguides to impart stable effective refractive index changes, while requiring only moderate (30 - 40 mW) power dissipation, avoiding excessive and potentially-damaging temperatures in the resistive heater material. Moreover, the localized hot spots generated by suspended heaters provide the possibility for independent trimming of multiple devices. Following thermal trimming with estimated temperatures of $\approx$ 300 - 400\textdegree C, maximum effective index changes of -5.18 $\times 10^{-3}$ and -7.9 $\times 10^{-4}$ were measured in SiN and Si waveguides, respectively. Thermally trimmed optical phase shifts were observed to be stable over a monitoring period of 23 days. Additionally, positive effective index changes up to $\approx$0.02 in SiN waveguides were observed at higher trimming temperatures ($\gtrsim$510\textdegree C), demonstrating bi-directional index trimming. The suspended heaters functioned as both trimming sections and efficient thermo-optic phase shifters, exhibiting $P_\pi = 1.7$ mW at $\lambda = 1550$ nm for SiN waveguides. Additional experiments and measurements were performed to study the underlying mechanism of the thermal trimming. From these results, we hypothesized that changes in the SiO$_2$ waveguide cladding with thermal treatment may have been a primary cause of the observed trimming effects at temperatures $\gtrsim$300\textdegree C, with changes in SiN waveguide cores occurring at $\gtrsim$510\textdegree C. We extrapolated the trimming data to 100 - 200\textdegree C operating temperatures over multiple years to estimate the thermal aging behavior of SiN waveguides in suspended heaters --- identifying the potential for significant drifting in their optical phase shift at $\gtrsim$150\textdegree C. Overall, the thermal trimming method demonstrated here offers a simple, flexible, and scalable route for the trimming of large photonic integrated circuits before/after packaging and without the introduction of additional fabrication and/or post-processing steps beyond those of standard Si photonics platforms. In addition, thermal trimming may provide possibilities for hardening Si photonic devices against thermal aging or counteracting optical phase shifts resulting from aging. With the thermal trimming and aging behavior being dependent on the properties of the SiO$_2$ waveguide cladding, SiN cores, and (potentially) passivation layer(s), future work is necessary to determine the variations in these effects across wafer lots and foundry processes. 

\appendix

\section*{Appendix A: Controlled phase trimming procedure}
\label{sec:AppendixA}

\begin{figure}[ht!]
    \centering
    \includegraphics[width=1\textwidth]{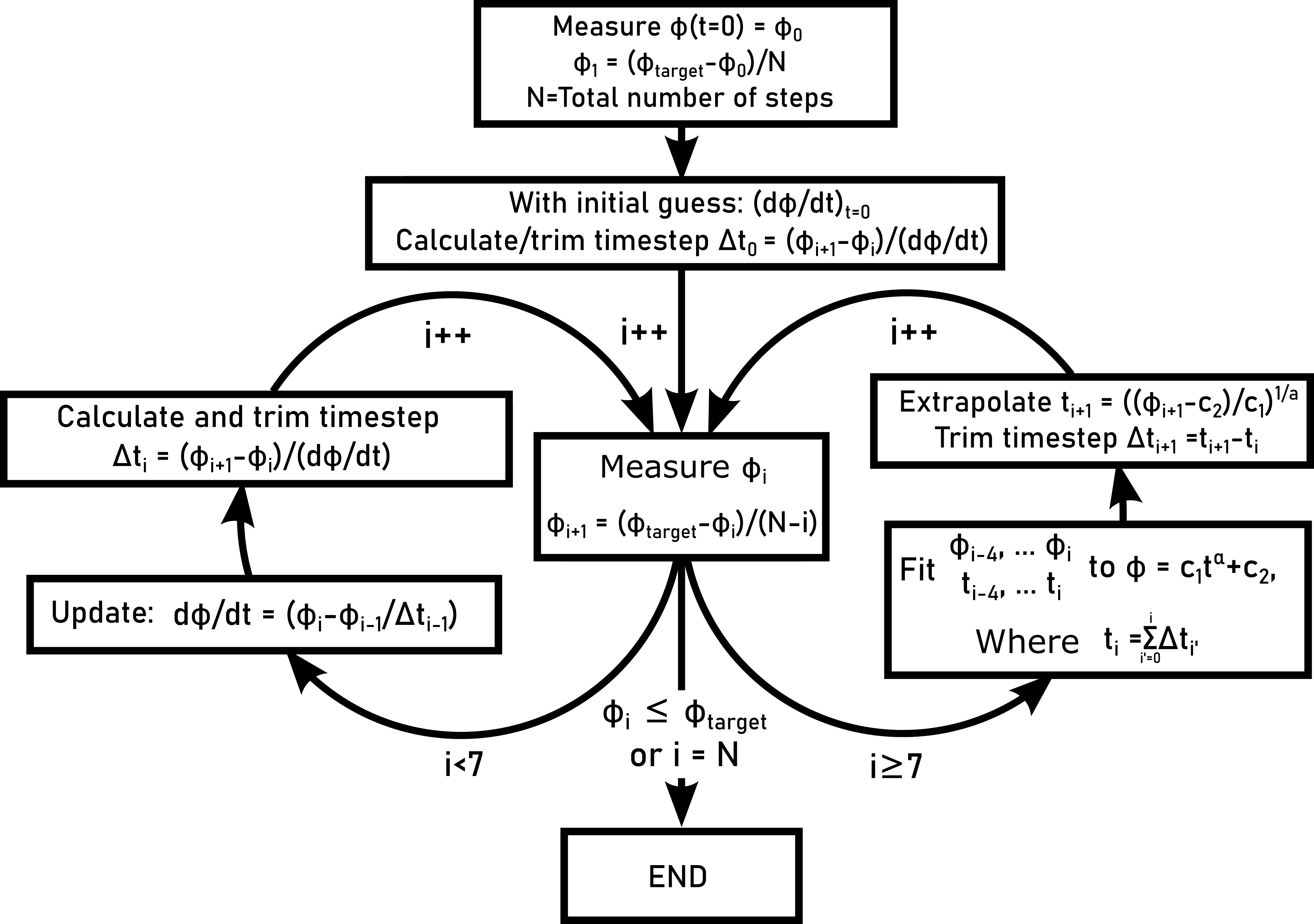}
            \caption{Block diagram illustrating the method of controlled trimming of the operating bias of the MZI switch in Fig. \ref{fig:bias}. The initial trimming time step ($\Delta t$) is calculated based on an initial guess of the trimming rate ($d\phi /dt$). The next 6 time steps are calculated based on the trimming rate of the previous trimming time step (left loop). After 7 trimming time steps, the subsequent trimming time steps are estimated by extrapolating a power law fit of the 5 previous trimming time steps (right loop) until the measured phase ($\phi$) reaches or passes the target phase ($\phi_{target}$).}
            \label{fig:diagram}
\end{figure}

Due to the observed recovery of a small fraction of the trimmed phase, it is necessary to overshoot the desired bias point during trimming in order to compensate.
We used this approach for trimming of the MZI switch in Fig. \ref{fig:bias}. The procedure is divided into two trimming sessions and outlined in Fig. \ref{fig:diagram}. 
The first trimming session (left loop in Fig. \ref{fig:diagram}) targeted some point between the initial bias and the target bias, and was followed by a monitoring period during which the recovery of the trimmed phase was used as an estimate of how much additional trimming was required. 
Next, for the second trimming session (right loop in Fig. \ref{fig:diagram}), the target bias was adjusted to trim beyond the desired bias point by the estimate obtained during the first trimming session.

To accurately predict the trimming duration, each session was split into steps spaced roughly 0.05 radians apart with $\phi$ being monitored at each step. 
The duration of the first thermal trimming session was determined by a preliminary guess for the trimming rate, and the subsequent 6 time steps were based on the updated estimate of the trimming rate derived from the preceding step. 
After 7 steps, we used the previous 5 data points to fit an observed power law dependence between the trimmed phase $\Delta \phi$ (at a trimming power of 30 mW) and the total trimming time $t$ of the form

\begin{equation}
\Delta\phi \propto t^{\alpha},
\end{equation}
to predict the subsequent time steps until the target was reached. The origin of this power law dependence is discussed in Appendix B. 
Figure \ref{fig:trim2bias} depicts the predicted and inferred $\phi$ at each time step during the trimming process of the device in Fig. \ref{fig:bias}. Accurate predictions of the required trimming durations are shown --- enabling precise adjustments to the target bias for mitigation of the partial post-trimming drift of $\phi$.

\begin{figure}[hb!]
    \centering
    \includegraphics[width=1\textwidth]{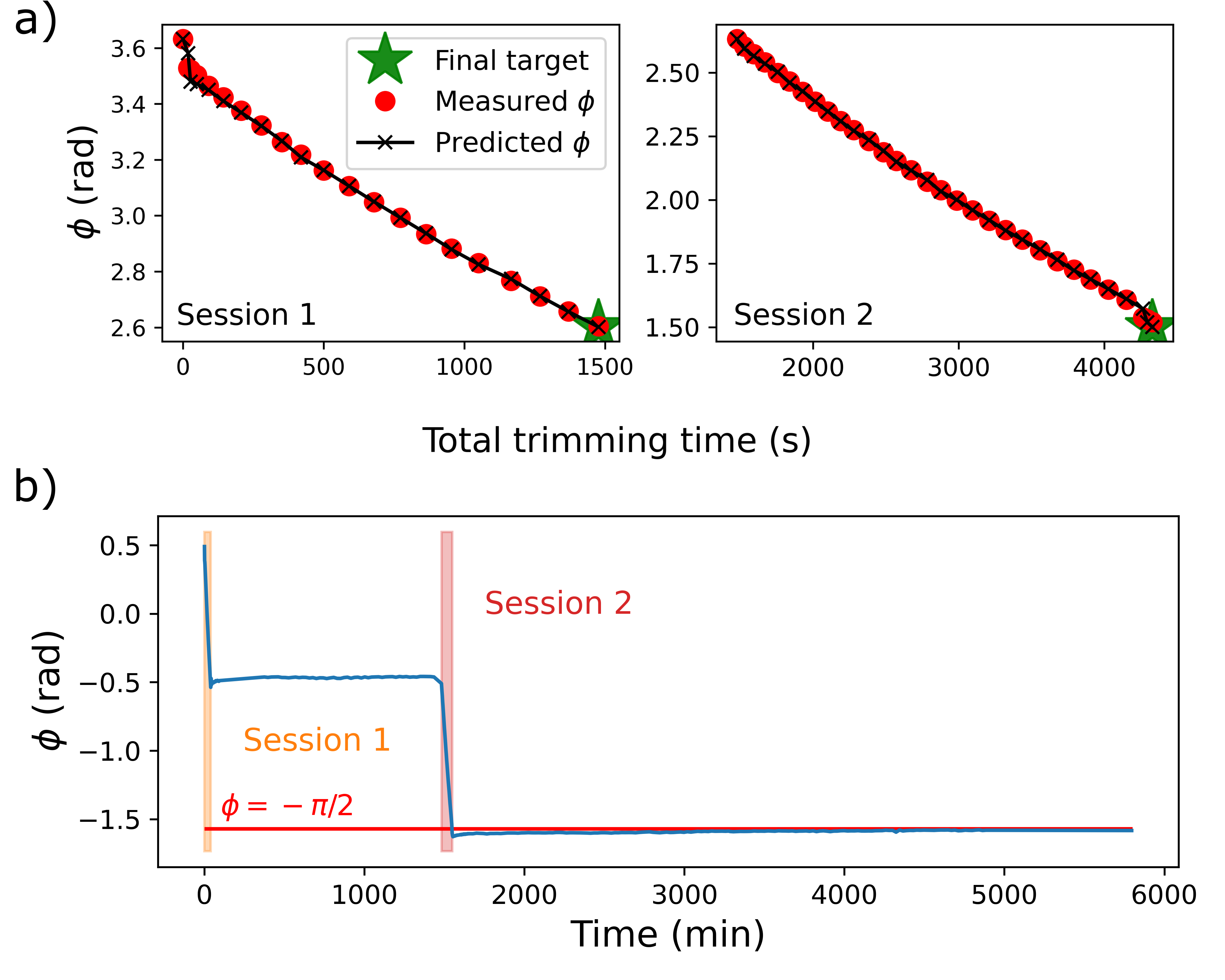}
            \caption{Visualization of the trimming of the MZI test device in Fig. \ref{fig:bias} to the target bias. (a) Measured and predicted phase difference between the two MZI arms ($\phi$) as a function of trimming time; measurements performed at Output 1. (b) Visualization of the two-step trimming process targeting $\phi=-\frac{\pi}{2}$.}
            \label{fig:trim2bias}
\end{figure}

\section*{Appendix B: Power-law relationship of phase shift due to thermal trimming}
\label{sec:AppendixB}

\begin{figure}[h!]
\centering
\includegraphics[width=1\textwidth]{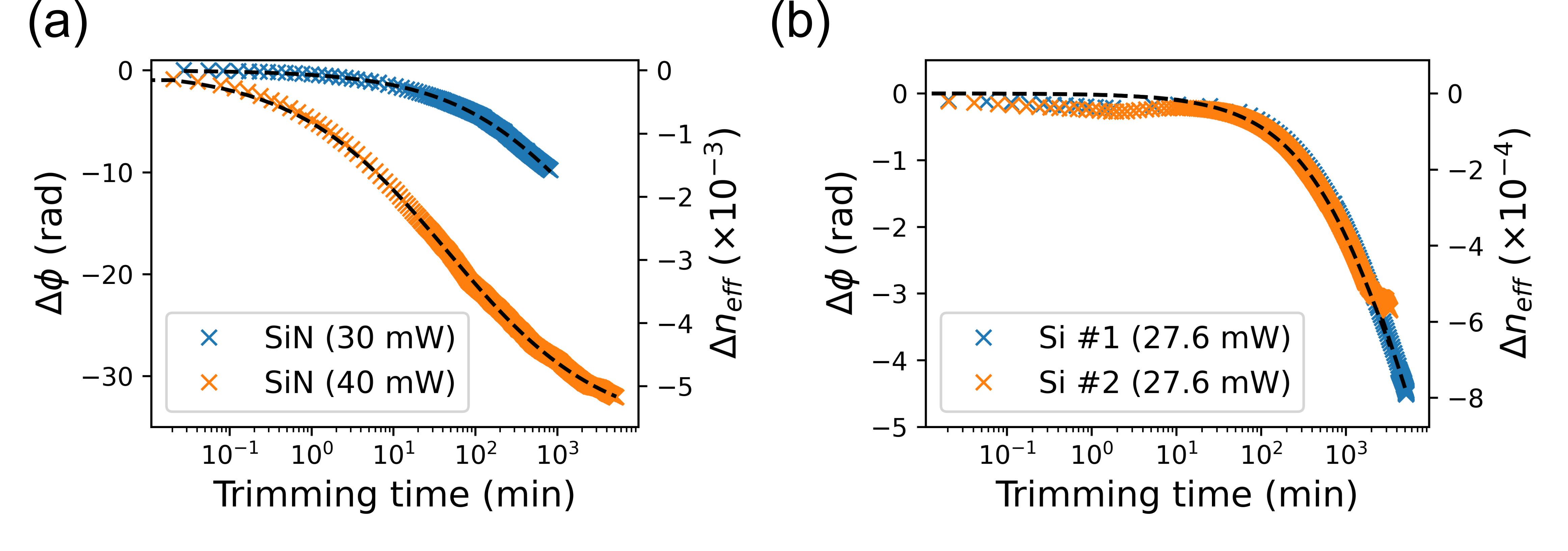}
\caption{
        Fitting of experimental (a) SiN trimming data from Fig. \ref{fig:trimming}(a) and (b) Si trimming data from Fig. \ref{fig:burn}(a) using Eq. \ref{eqn:powerlaw2}. 
        The fitting parameters for $\Delta \phi$ vs. trimming time are summarized in Table \ref{table:fitting_parameters}.}
            \label{fig:fit}
\end{figure}

The magnitude of the trimmed phase due to thermal annealing empirically fits to a modified version of the function proposed in Ref. \citenum{Erdogan1994} (to describe the thermal decay of the reflectivity of ultraviolet-written fiber Bragg gratings \cite{Erdogan1994,Bertrand2015,Kharakhordin:19}):
\begin{equation}
\label{eqn:powerlaw2}
\Delta\phi = B\left(\frac{At^{\alpha}}{1+At^\alpha}\right),
\end{equation}
where $B$, $A$ and $\alpha$ are fitting parameters. Fits of the trimming data presented in Figs. \ref{fig:trimming}(a) and \ref{fig:burn}(a) using Eq. \ref{eqn:powerlaw2} are shown in Fig. \ref{fig:fit}.

In a fitting window where the variation of the term $At^\alpha$ is much smaller than 1, Eq. \ref{eqn:powerlaw2} can be approximated as a simple power-law relationship of the form 
\begin{equation}
\Delta\phi \approx  c_1 t^\alpha,
\end{equation}
with fitting parameters $c_1$ and $\alpha$. This power-law relationship was used for controlled trimming of the MZI switch in Fig. \ref{fig:bias} (see Appendix A). 

\clearpage

\begin{table}[h!]
\caption{$\Delta \phi$ fitting parameters corresponding to Fig. \ref{fig:fit}.} 
    \centering
    \begin{tabular}{c | c | c | c | c }
    \hline
         \begin{tabular}{@{}c} Sample \end{tabular} &  
        \begin{tabular}{@{}c} $P_{Appl}$ (mW) \end{tabular} &
        \begin{tabular}{@{}c} $A$ \end{tabular} &
        \begin{tabular}{@{}c} $B$ \end{tabular} &
        \begin{tabular}{@{}c} $\alpha$ \end{tabular}\\[0.5ex] 
    \hline
    SiN & 30 & 0.0167 & 27.9 & 0.523 \\
    \hline
    SiN & 40 & 0.171 & 35.6 & 0.464 \\
    \hline
    Si \#1 & 27.6 & 0.00183 & 9.79 & 0.717 \\
    \hline 
    Si \#2 & 27.6 & 0.00224 & 8.45 & 0.727 \\
    \hline
    \end{tabular}
    \label{table:fitting_parameters}
\end{table}

\section*{Appendix C: Variation of initial relative optical phase shift between MZI arms}
\label{sec:AppendixC}

\begin{figure}[ht!]
    \centering
    \includegraphics[width=0.6\textwidth]{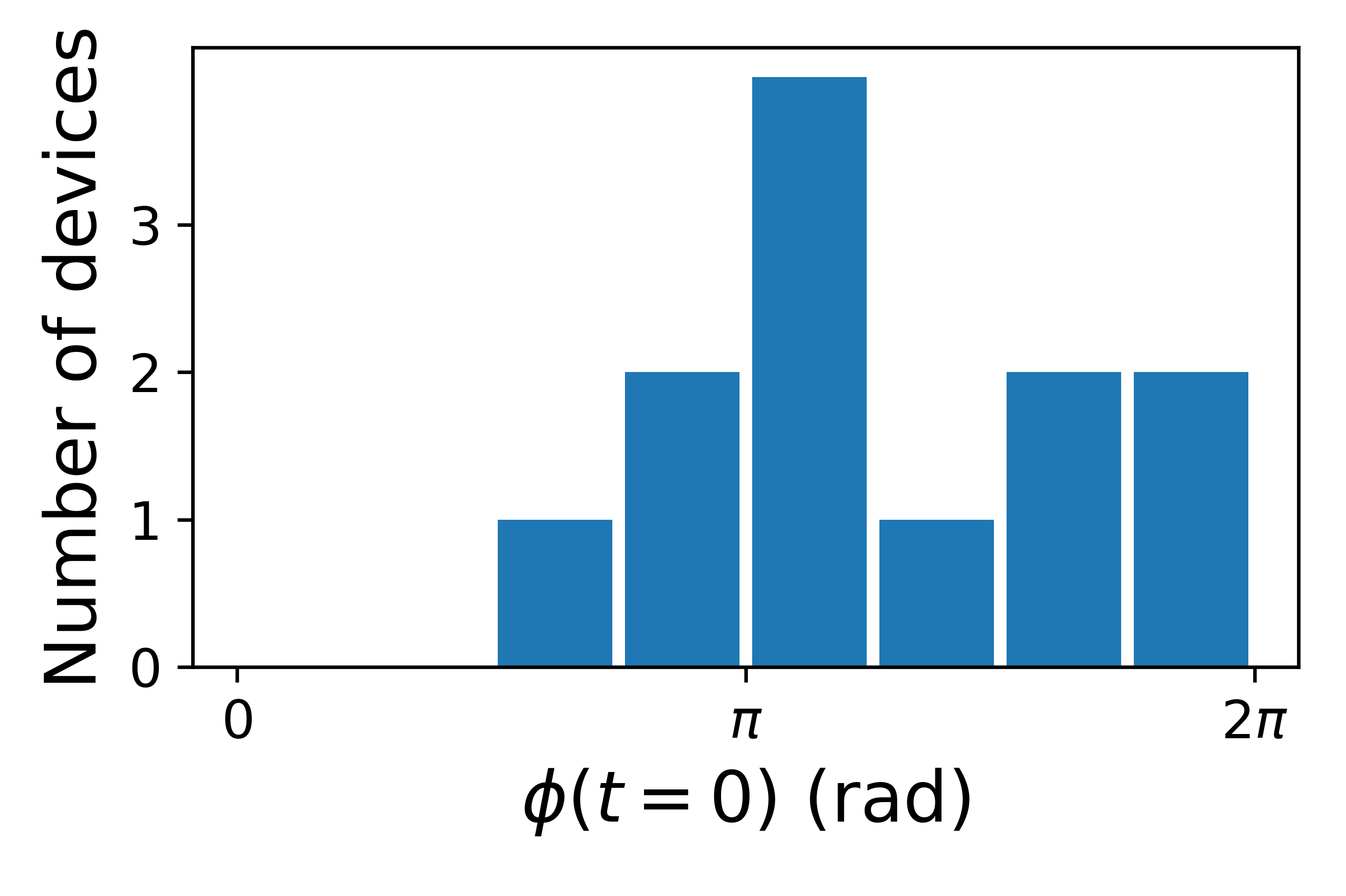}
            \caption{Histogram of initial relative optical phase shift ($\phi$) between the 2 MZI arms of the SiN test device across 12 chips.}
            \label{fig:init}
\end{figure}

\begin{backmatter}
\bmsection{Funding}
Max-Planck-Gesellschaft
\bmsection{Acknowledgments}
This work was supported by the Max Planck Society. The authors acknowledge the SiEPIC program, CMC Microsystems, and Advanced Micro Foundry for the fabrication of the silicon photonic devices.
\bmsection{Disclosures}
The authors declare no conflicts of interest.
\bmsection{Data availability}
Data underlying the results presented in this paper are not publicly available at this time but may be obtained from the authors upon reasonable request.
\end{backmatter}

\end{document}